\documentclass{nature2}
\usepackage{graphicx}
\usepackage[usenames]{color}
\usepackage{xcolor}
\usepackage[export]{adjustbox}

\usepackage{amsmath}
\usepackage{amssymb}

\title{Rules governing Jahn--Teller order in\\ Prussian blue analogues}

\author{Elodie A. Harbourne,$^1$ John Cattermull,$^{1,2}$ Hanna L. B. Boström,$^{1,3,4}$ Rebecca Witte,$^{1,2}$ Nikolaj Roth,$^{1}$ Mauro Pasta,$^2$ David A. Keen$^{5}$ and Andrew L. Goodwin$^{1\ast}$}

\begin{document}

\maketitle

\begin{affiliations}
 \item Inorganic Chemistry Laboratory, University of Oxford, Oxford OX1 3QR, U.K.
 \item Department of Materials, University of Oxford, Oxford OX1 3PH, U.K.
 \item Wallenberg Initiative Materials Science for Sustainability, Department of Materials and Environmental Chemistry, Stockholm
University, SE-114 18 Stockholm, Sweden
 \item Department of Chemistry, Stockholm
University, SE-114 18 Stockholm, Sweden
 \item ISIS Facility, Rutherford Appleton Laboratory, Harwell Campus, Didcot OX11 0QX, U.K.
\end{affiliations}

\begin{abstract}

Jahn--Teller distortions of transition-metal coordination environments link orbital occupancies to structure. In the solid state, such distortions can be strongly correlated through the propagation of strain and/or through orbital interactions. Cooperative Jahn--Teller (CJT) order of this kind affects the electronic, magnetic, and structural properties of the materials in which it occurs. Conventionally studied in dense ceramics, CJT order also occurs in hybrid materials, albeit the underlying phenomenology is not well established. Here we use synchrotron powder X-ray diffraction measurements to identify the compositional factors governing cooperative Jahn–Teller order in a series of Prussian blue analogue (PBA) families. We develop a simple microscopic model based on the dual considerations of strain and crystal-field stabilisation that rationalises the stability, extent of CJT order, and crystallite strain measured experimentally. This model shows how PBA compositions might be tuned to control the emergence and nature of CJT effects, and predicts universal phase behaviour for JT-active PBAs more generally. Our results establish a microscopic framework for understanding and controlling CJT effects in PBAs, and reveal an interplay between compositional, structural, and orbital degrees of freedom closely analogous to that of the manganite perovskites.

\end{abstract}

\section*{Introduction}

Cooperative Jahn--Teller (CJT) effects play a number of important roles---sometimes useful and sometimes disadvantageous---in the physical behaviour of very different classes of solid materials.\cite{Gehring_1975,Goodenough_1998,Bersuker_2021} In the manganite perovskites, to take one positive example, CJT instabilities compete with charge, magnetic, and tilt order to drive the useful property of colossal magnetoresistance.\cite{Coey_1999,Khomskii_2003,Goodenough_2004,Hotta_2000} Likewise, dynamic CJT effects are implicated both in the strong anharmonicity of chalcogenide thermoelectrics\cite{Wang_2024b} and in the pairing mechanisms of high-temperature superconductors.\cite{Han_2003,Sergeeva_2004} By contrast, CJT-driven symmetry-breaking is generally an undesirable process in cathode materials, such as Li$_x$Mn$_2$O$_4$ and Li$_x$NiO$_2$,\cite{Englman_1970,Thackeray_1983,Yamada_1999,GenreithSchriever_2024} where switching on and off cooperative distortions during battery cycling results in gradual mechanical failure and capacity fade.\cite{Yamada_1995,Aurbach_1999,Kalyani_2005} It is no surprise therefore that establishing rules governing CJT order in solids---both empirical and theoretical---has been a crucial aspect of functional materials design.\cite{Halcrow_2013,Bersuker_2021b} The development of such rules is arguably most mature in the context of dense transition-metal ceramics, such as perovskites, spinels, and rocksalts. In these systems, two primary mechanisms dominate. The first is vibronic coupling, in which the orbital occupation of one JT-active site couples with that of others through lattice strain.\cite{Bersuker_2021} And the second is the Kugel--Khomskii mechanism, whereby electronic exchange drives CJT order directly through orbital interactions alone.\cite{Kugel_1982,Khomskii_2021}

\begin{figure}
\begin{center} 
\includegraphics{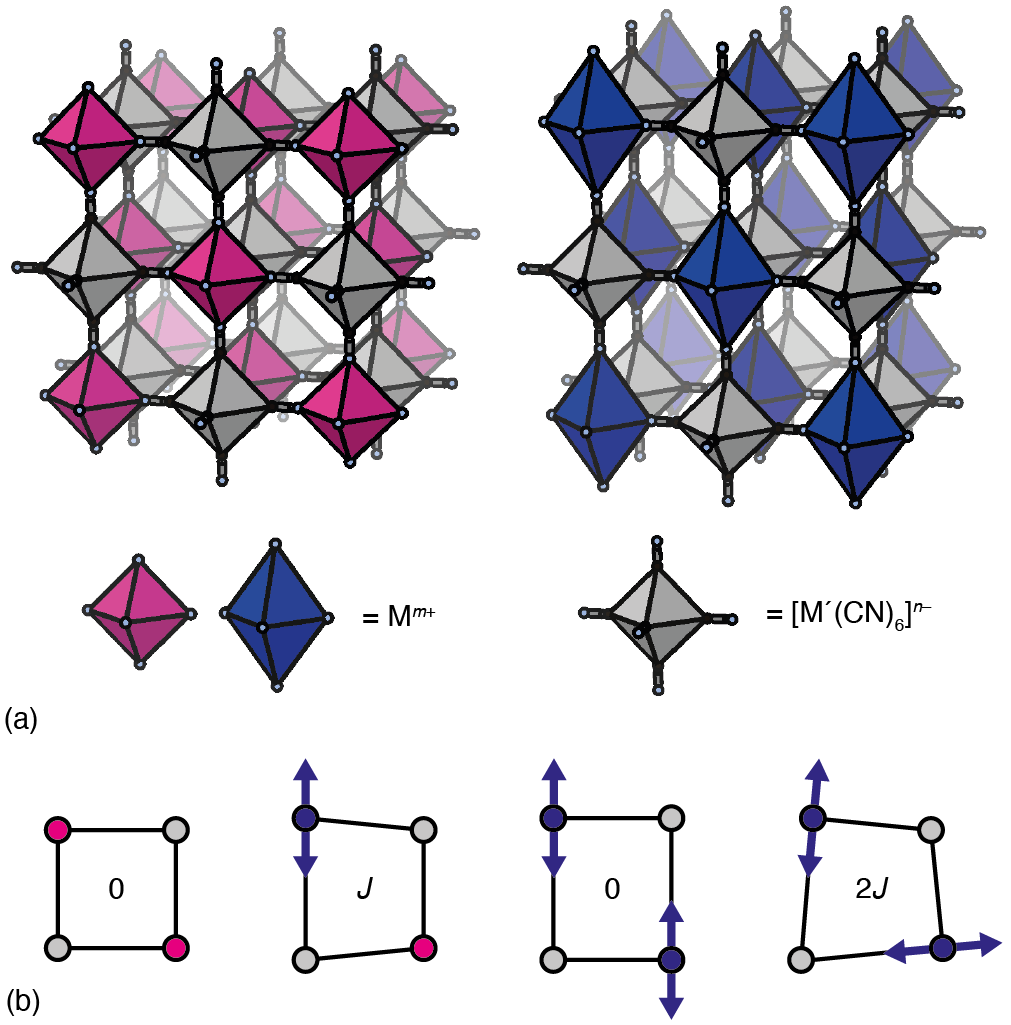}
\end{center}
\caption{\label{fig1}\footnotesize{\bf Structure and CJT effects in PBAs.}  (a) The basic PBA structure type involves alternating M$^{m+}$ (magenta/blue) and [M$^\prime$(CN)$_6$]$^{n-}$ (grey) octahedral ions arranged on a simple cubic lattice. Counterions needed for charge balance and/or solvent molecules can occupy the cube cavities and are not shown here. The aristotype (left) has cubic symmetry. In systems such as Cu[Pt(CN)$_6$], the JT effect drives axial elongation of the M-site Cu$^{2+}$ coordination polyhedra. This local distortion is known to correlate when all M-site ions are JT active and there are no M$^\prime$-site vacancies, reducing the crystal symmetry from cubic to tetragonal (right). (b) Strain model for cooperative JT effects in PBAs. Squares represent windows of the PBA structure; \emph{i.e.}\ four-membered rings of connected octahedra perpendicular to the parent $\langle100\rangle$ crystallographic directions. Coloured circles represent M/M$^\prime$ sites: magenta = JT-inactive M-site or JT-active M-site with out-of-plane axial elongation, blue = JT-active M-site with in-plane elongation, grey = M$^\prime$-site. Blue arrows denote the local axis of JT-driven elongation. The four possible scenarios give rise to strain energies that are approximate multiples of a common energy scale $J$.}
\end{figure}

Hybrid framework materials provide an intriguing contrast because both conventional mechanisms for CJT order are expected to weaken:\cite{Li_2017} the open and mechanically compliant architectures of hybrids reduce elastic coupling,\cite{Tan_2011} while increased metal-metal separations suppress direct orbital interactions.\cite{Kostakis_2010} Nevertheless CJT distortions do occur,\cite{Stroppa_2011,Bostrom_2018,Cattermull_2021} and here we are particularly concerned with their phenomenology amongst Prussian blue analogues (PBAs)---an especially important and compositionally diverse class of hybrid frameworks with applications spanning a variety of very different fields, \emph{e.g.}\ energy storage, catalysis, ion conduction, photoresponse, and magnetism.\cite{Kaye_2005,Ohkoshi_2010,Ferlay_1995,Bleuzen_2000,LeKhac_1998,Wessells_2011} The basic PBA structure type is shown in Fig.~\ref{fig1}(a): M$^{m+}$ and [M$^\prime$(CN)$_6$]$^{n-}$ ions alternate on the nodes of a simple cubic lattice. It is common for PBAs to exhibit [M$^\prime$(CN)$_6$]$^{n-}$ vacancies, and additional cations can be incorporated within the cubic framework cavities as needed for charge balance.\cite{Cattermull_2021} A simple example of CJT order in this family is given by the material Cu[Pt(CN)$_6$], the structure of which has tetragonal symmetry to accommodate the cooperative elongation of Cu$^{2+}$ coordination environments along a single common axis [Fig.~\ref{fig1}(a)].\cite{Siebert_1975,Chapman_2006b} A similar distortion is triggered on extraction of K$^+$ ions from the initially-undistorted cathode material KMn[Fe(CN)$_6$] as Mn$^{2+}$ is oxidised to Jahn--Teller-active Mn$^{3+}$.\cite{Bie_2017} The strain induced in this process is considered key in limiting the reversible electrochemistry of the system.\cite{Jiang_2019,Cattermull_2026} Varying [M$^\prime$(CN)$_6$]$^{n-}$ vacancy fraction must also switch on and off CJT order, because the nonstoichiometric cathode material K$_{0.71}$Cu[Fe(CN)$_6$]$_{0.72}\cdot$3.7H$_2$O---which exhibits vastly superior cycling behaviour to KMn[Fe(CN)$_6$]---has no cooperative distortion despite its high Cu$^{2+}$ content.\cite{Wessells_2011} So, asking what rules might govern CJT order in this family is an important open question of direct relevance to many of the practical applications of functional PBA materials.\cite{Maiti_2026}

Here we address this question by studying three compositional series---\emph{viz}.\ Mn$_{1-x}$Cu$_x$[Pt(CN)$_6$], Cs$_x$Cu[Fe(CN)$_6$]$_{(2+x)/3}$, and Mn$_{1-x}$Cu$_x$[Co(CN)$_6$]$_{2/3}$---using high-resolution synchrotron X-ray diffraction measurements to characterise their phase stability and extent of long-range CJT order as a function of JT dilution and hexacyanometallate vacancy concentration. We have chosen these three particular series because they allow us to identify the separate effects of (i) increasing JT concentration in a vacancy-free system, (ii) increasing vacancy concentration in a fully JT-active system, and (iii) introducing JT activity within a partially-vacant PBA. We infer from our experimental measurements of the first two families a simple microscopic model governing CJT order based on the competition between strain and crystal-field effects. This model allows us to develop a generic phase diagram for PBAs, identifying compositions for which CJT-driven symmetry breaking is expected. We demonstrate the veracity of the model by comparing prediction against observation for the third family. A key result is that local CJT order in vacancy-rich PBAs adopts different patterns to the long-range order observed at low vacancy concentrations, suggesting that the family harbours its own complex interplay between compositional, structural, and orbital degrees of freedom not unlike that found in the manganite perovskites.\cite{Goodenough_1955,Millis_1998,Coey_1999,Tokura_2000}

\section*{Results and Discussion}

\begin{figure}
\begin{center} 
\includegraphics{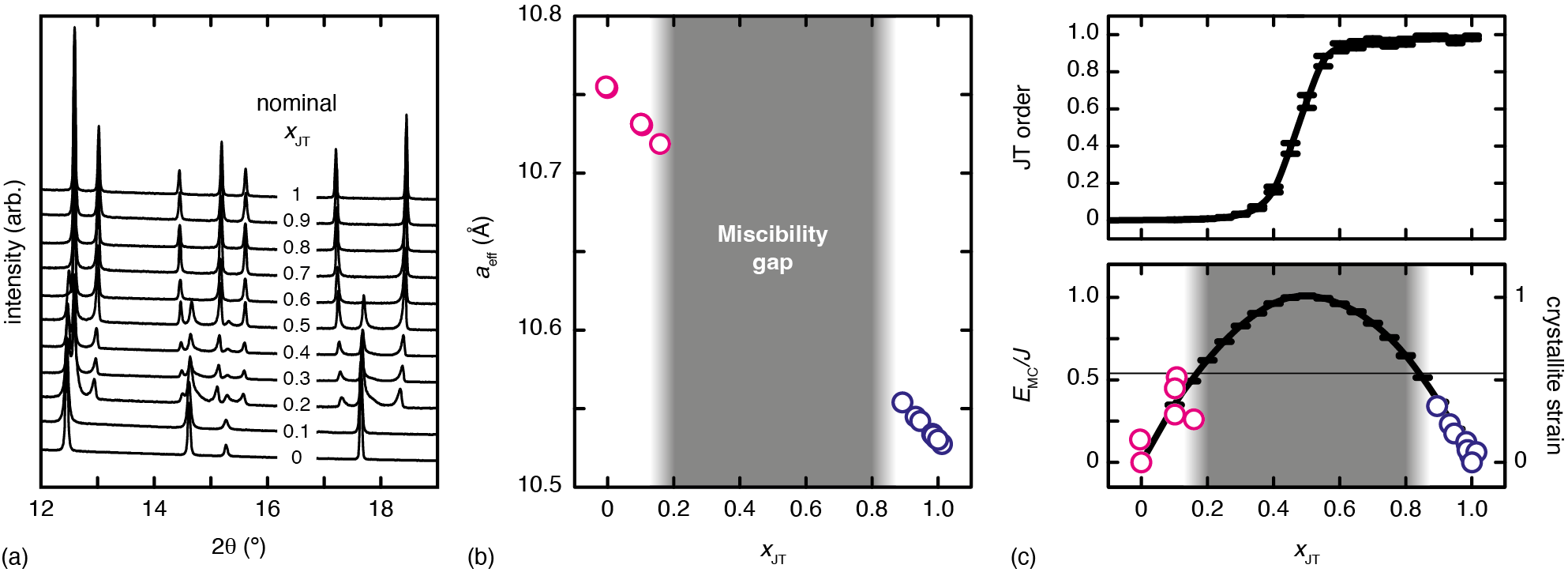}
\end{center}
\caption{\label{fig2}\footnotesize{\bf Structural behaviour of Mn$_{\textbf 1\boldsymbol{-x}}$Cu$_{\boldsymbol x}$[Pt(CN)$_{\textbf 6}$] PBAs.}  (a) Representative region of the X-ray diffraction patterns ($\lambda = 0.82507$\,\AA) measured as a function of nominal composition. On substitution of Mn$^{2+}$ by Cu$^{2+}$ (increasing $x_{\rm JT}$, bottom--top) the simple diffraction pattern of the cubic aristotype gradually decreases in intensity, and a more complex pattern associated with the tetragonal CJT structure type emerges. Small shifts in peak positions and variation in peak widths are also observed. (b) The effective lattice parameters $a_{\rm eff}=\sqrt[3]{4V/Z}$ extracted from the data in (a) using Rietveld refinement, and the corresponding JT compositions $x_{\rm JT}$ estimated from Vegard's law. Data points corresponding to cubic (undistorted) phases are shown in magenta, and those of tetragonal (distorted) phases are shown in blue. These data show the existence of a miscibility gap. (c) Coarse-grained MC simulations driven by the strain model shown in Fig.~\ref{fig1}(b) give ground-state structures with varying degrees of CJT order (top) and residual strain (bottom, solid line). The isotropic strain broadening measured in our Rietveld refinements of the data in panel (a) are shown as open circles, and are coloured according to the presence (blue) or absence (magenta) of long-range JT order. The thin horizontal line denotes the accessible thermal energy at room temperature for $J=5$\,kJ\,mol$^{-1}$ as discussed in the main text. Standard uncertainties in the data shown in panels (b) and (c) are smaller than the symbols; estimated uncertainties in MC-derived quantities are shown as error bars.}
\end{figure}

We first investigated the solid-solution Mn$_{1-x}$Cu$_x$[Pt(CN)$_6$]. The Mn$^{2+}$ and Cu$^{2+}$ endmembers gave X-ray diffraction patterns characteristic of the cubic $Fm\bar3m$ and tetragonal $I4/mmm$ PBA phases anticipated for JT-free and CJT-ordered states, respectively.\cite{Chapman_2006b} For intermediate compositions, we observed varying fractions of both cubic and tetragonal phases, with lattice parameters slightly different to those of the corresponding endmembers [Fig.~\ref{fig2}(a)]. We estimated the composition of each phase assuming a Vegard's law relation with respect to cell volume, and determined phase fractions using constrained Rietveld refinement. There was good agreement between overall compositions determined crystallographically and from atomic absorption spectroscopy measurements. Crucially, our data indicated a miscibility gap for $0.15\lesssim x\lesssim0.85$ characteristic of limited solubility of JT-active ions in an undistorted crystal lattice and \emph{vice versa} [Fig.~\ref{fig2}(b)]. Samples with nominal compositions within this `forbidden' range formed mixtures of cubic and tetragonal phases whose compositions differed only slightly from those of the corresponding endmembers. Incorporating some JT-active Cu within the cubic Mn structure (or JT-inactive Mn within the tetragonal Cu structure) clearly led to strong crystallographic strain, as reflected in the widths of Bragg reflections. Our Rietveld refinements allowed us to quantify this strain, which we found to vary symmetrically with composition [Fig.~\ref{fig2}(c)].

Anisotropic strain models are often used to rationalise CJT effects in many conventional oxide perovskites,\cite{Halperin_1971,Gehring_1975} and so we sought to establish the simplest such model that could explain both the miscibility gap and compositional strain in this first PBA family. Rather than attempting a full atomistic description (which might not be transferable), our focus is on coarse graining to capture the phenomenology of cooperative JT order. The model we developed was based on deformations of rectangular windows of the PBA framework as a function of the number and (relative) orientation of long JT-active bonds. For each window, there are four possible scenarios [Fig.~\ref{fig1}(b)], and simple geometric considerations give strain energies as integral multiples of a single characteristic strain-energy scale, $J$. Direct first-principles validation of this energy scale is impractical because of the size and configurational complexity of JT-dilute PBA supercells, but the elastic moduli suggest $J\sim5$\,kJ\,mol$^{-1}$ (see SI 3.1).\cite{Felix_2018} We used this simplified strain model as the basis of a series of Monte Carlo (MC) simulations in which Cu/Mn sites were allocated randomly and JT-orientations were treated as free variables. For compositions $x\gtrsim0.4$ the model exhibited a phase transition to a CJT-ordered ground state at low effective temperatures [Fig.~\ref{fig2}(c)]. The ground-state energies varied symmetrically with composition; these energies measure both the enthalpic cost of solid-solution formation and the residual strain arising from elastic mismatch between short- and long-range CJT effects within the mixed-composition phase. The miscibility gap in Mn$_{1-x}$Cu$_x$[Pt(CN)$_6$] emerges because the available thermal energy during synthesis is insufficient to access the high-energy intermediate compositions. Likewise the experimentally-observed composition--strain relationship is captured naturally by this MC model [Fig.~\ref{fig2}(c)]. An advantage of the model being so simple is that it generalises. Across different PBA families, variations in the corresponding elastic tensors will change $J$, broadening or narrowing the miscibility gap without altering the qualitative phase behaviour. Indeed this same microscopic picture helps rationalise the nonequilibrium transformation pathways of K$_x$Mn[Fe(CN)$_6$] cathodes, where electrochemical cycling changes the fraction of JT-active Mn$^{3+}$ centres.\cite{Cattermull_2026}

\begin{figure}
\begin{center} 
\includegraphics{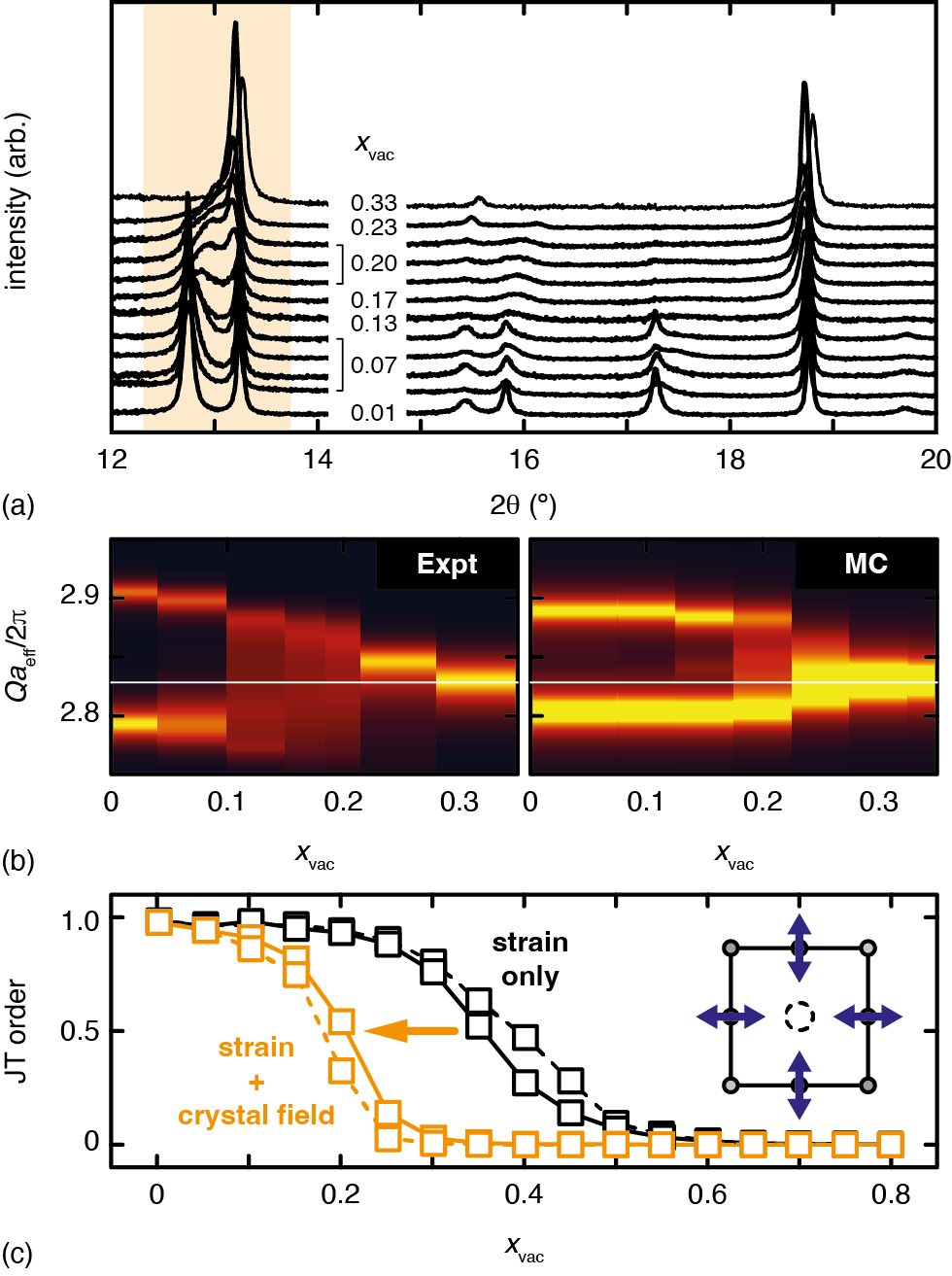}
\end{center}
\caption{\label{fig3}\footnotesize{\bf Vacancy-driven CJT melting in Cs$_{\boldsymbol x}$Cu[Fe(CN)$_{\mathbf 6}$]$_{\mathbf{(2}\boldsymbol{+x}\mathbf {)/3}}$.}  (a) Representative X-ray diffraction patterns ($\lambda=0.826060$\,\AA) measured as a function of vacancy fraction. (b) The region of the X-ray diffraction patterns of Cs$_x$Cu[Fe(CN)$_6$]$_{(2+x)/3}$ spanning the $\{220\}$ reflections, which are split in the tetragonal CJT-ordered phase but not in the cubic JT-disordered phase (yellow shaded region in panel (a)). The thin horizontal line at $2\sqrt{2}$ reciprocal lattice units indicates the high-symmetry $\{220\}$ position. The left-hand panel shows experimental data: melting of CJT order is observed around $x_{\rm vac}\simeq0.2$, with strong broadening at critical compositions. The right-hand panel shows the results from MC simulations driven by a combined strain + crystal-field model. (c) Dependence of ground-state CJT order on M$^\prime$-site vacancy concentration determined using MC simulations. The black data points correspond to simulations driven only by the strain model illustrated in Fig.~\ref{fig1}(b); the yellow data points correspond to the model which also takes into account the crystal-field preference for JT orientations to point towards vacancies (see inset). Data connected by solid lines correspond to random vacancy distributions, and data connected by dashed lines correspond to configurations in which vacancies were distributed with the correlations determined for single-crystal samples in Ref.~\citenum{Simonov_2020}. Standard uncertainties in the data shown in panel (c) are smaller than the symbols.}
\end{figure}

We next considered the effect of incorporating hexacyanometallate vacancies on the emergence of CJT order. Experimentally, we explored this point by characterising the diffraction behaviour of the Cs$_x$Cu[Fe(CN)$_6$]$_{(2+x)/3}$ PBA family, for which the concentration of JT-active Cu$^{2+}$ is constant and maximal [Fig.~\ref{fig3}(a)]. As expected from the arguments presented above, the vacancy-free material CsCu[Fe(CN)$_6$] is tetragonally distorted (space group $I\bar4m2$), reflecting long-range CJT order of the Cu$^{2+}$ ions.\cite{Svensson_2019} On increasing the [Fe(CN)$_6$]$^{3-}$ vacancy fraction $x_{\rm vac}=(1-x)/3$, CJT order eventually disappears such that the limiting composition Cu[Fe(CN)$_6$]$_{2/3}$ adopts a distortion-free cubic structure with average symmetry $Fm\bar3m$. The critical vacancy fraction for CJT order was $x_{\rm vac}\simeq20$\%, around which the Bragg reflections in the X-ray diffraction patterns showed particularly strong anisotropic peak broadening [Fig.~\ref{fig3}(a)]. This broadening likely involves contributions from lattice strain, some compositional variations amongst crystallites, and also diffuse scattering as observed elsewhere at morphotropic phase boundaries.\cite{Zhang_2014,Zhang_2018}

Arguably the simplest way in which to incorporate the effect of vacancies within a strain model is to assume that the presence of a vacancy relieves all strain in the neighbouring windows. Such a modification was easily implemented within our MC simulations, and we found that ground-state CJT order indeed vanished on increasing vacancy concentration. The critical vacancy fraction in these MC simulations, however, was $x_{\rm vac}\simeq0.5$---such that CJT order was more resilient in this model than observed in practice [Fig.~\ref{fig3}(c)]. We considered the possibility that this discrepancy might reflect the importance of vacancy correlations (\emph{e.g.}\ as established in Ref.~\citenum{Simonov_2020}), but found similar behaviour in our MC simulations for both random- and correlated-vacancy distributions [Fig.~\ref{fig3}(c)]. The discrepancy between experiment and model suggested that vacancies do more than to relieve elastic strain alone.

The obvious ingredient missing in our microscopic model was to consider how vacancies might influence JT-axis orientations of neighbouring sites. It is known from the layered structure of CuNi(CN)$_4$,\cite{Chippindale_2015} for example, that the fully-occupied $e_g^\ast$ orbital of Cu$^{2+}$ is always oriented orthogonal to its planar N$_4$ coordination environment---\emph{i.e.}\ towards the vacancies in its coordination shell. Likewise, in a recent DFT study of copper hexacyanoferrates,\cite{Wang_2024} the JT axes of Cu$^{2+}$ ions were found to point towards vacancies, irrespective of whether or not the relevant coordination site was occupied by water. This anisotropy is ultimately a crystal-field effect: both H$_2$O and vacancies are weaker-field ligands than N-bound cyanide.\cite{Figgis_2000} Orienting the local elongation along this weaker-field direction reduces the energy penalty of double occupation of the corresponding antibonding $e_g^\ast$ orbital. Consequently, we introduced an additional crystal-field term in our MC Hamiltonian that penalised JT orientations pointing towards an occupied site [Fig.~\ref{fig3}(c)]. Once again the energy scale of this crystal field term $J^\prime$ is difficult to calculate from \emph{ab initio} methods for these compositionally-complex systems; however, irrespective of the exact value, MC simulations incorporating this additional crystal-field term always showed a lower critical vacancy fraction for cooperative Jahn--Teller order. In fact the simple approximation $J^\prime/J\sim1$ gives behaviour surprisingly close to experiment for Cs$_x$Cu[Fe(CN)$_6$]$_{(2+x)/3}$ [Fig.~\ref{fig3}(b,c)], with the result again independent of vacancy correlations. Our simulations also reproduced qualitatively the strain observed at the order/disorder transition. Varying the relative strengths of crystal field and strain terms---\emph{i.e.}\ the $J^\prime/J$ ratio---simply shifted the critical vacancy fraction for CJT order, and so we might expect this threshold to differ slightly amongst different PBA families. In any case, since the preferred axis for JT elongation necessarily rotates around a vacant site [Fig.~\ref{fig3}(c)], the universal effect of crystal-field considerations is to frustrate the coaxial alignment favoured by strain and responsible for CJT order.

\begin{figure}
\begin{center} 
\includegraphics{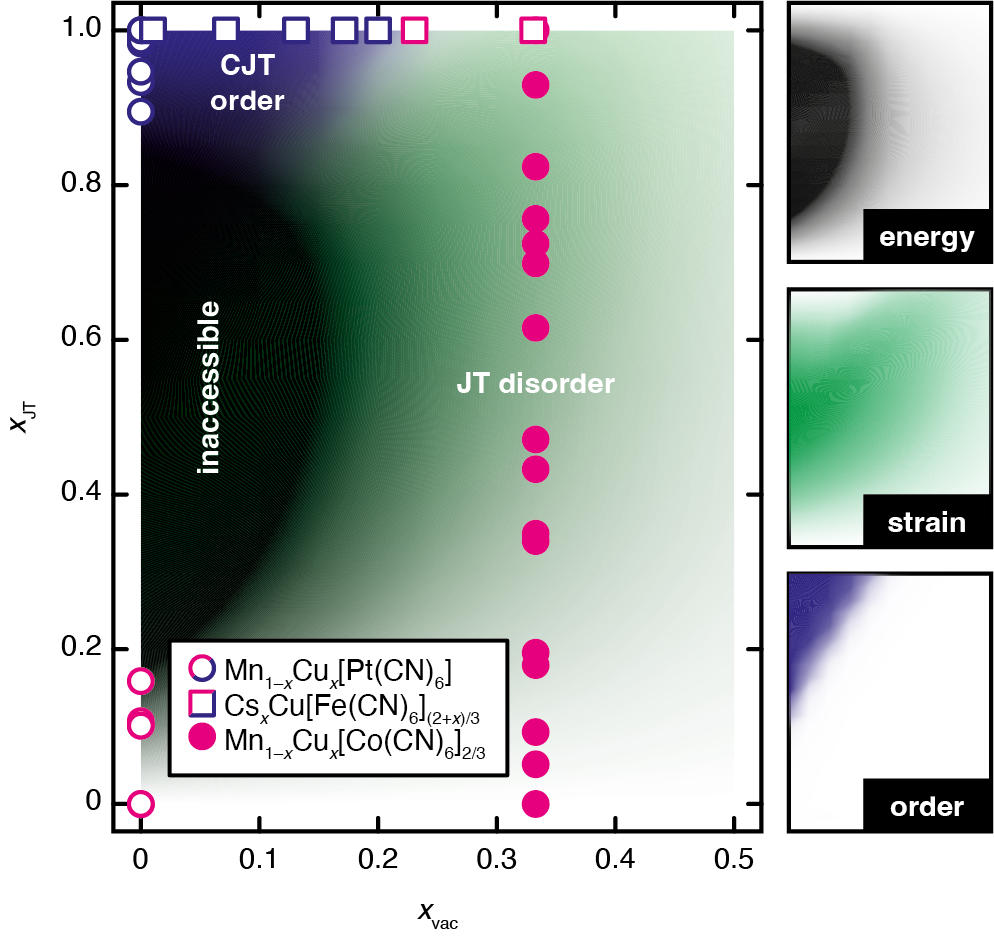}
\end{center}
\caption{\label{fig4}\footnotesize{\bf Ground-state phase map for PBAs arising from the competition between strain and crystal-field considerations.} MC simulations driven by strain and crystal-field terms give ground-state structures whose energies, residual strain, and degree of CJT order vary systematically as a function of M-site JT activity and [M$^\prime$(CN)$_6$]$^{n-}$ vacancy concentration. The single parameter of this model, $J$, determines the scales of the energy and strain terms, but affects neither the emergence of CJT order nor the basic form of the features shown here. The right-hand panels show the individual contributions of energy, strain, and CJT order to the main phase map. CJT order is expected only for PBA samples with high fractions of JT-active M-sites and low M$^\prime$ vacancy concentrations (blue region in top-left). Low-vacancy phases with intermediate $x_{\rm JT}$ fractions are unstable with respect to segregation into JT-poor and JT-rich phases, giving rise to an inaccessible region of phase space (coloured black). The boundary of this region is sensitive to the magnitude of $J$, which may vary from PBA family to PBA family. Likewise the relative strengths of strain and crystal-field effects determine the vacancy fraction at which CJT order emerges (\emph{i.e.}\ extent of the blue region), which again may vary amongst PBA families. The phase map is dominated by a continuous region of JT-disordered states, many with large residual strains (shown in green). JT orientations are not random in these states, but are governed by the competing effects of strain (which favours JT coalignment) and crystal-field (which favours reorientation of JT axes towards vacancies). The coordinates corresponding to each of the systems for which we have experimental X-ray diffraction measurements are shown as data points coloured magenta if the corresponding structure is cubic (no CJT order) or blue if tetragonal (CJT order).}
\end{figure}

Together, the ingredients of strain and crystal-field define a simple microscopic model from which a general ground-state phase diagram for JT-active PBAs follows [Fig.~\ref{fig4}]. The phase behaviour emerges by considering the interplay of ground-state energy (governing `forbidden' regions), residual strain, and cooperative order. One key (and testable) prediction of our model is that vacancies stabilise solid solutions of partial JT activity, such that the miscibility gap observed for stoichiometric PBAs (\emph{e.g.}\ the hexacyanoplatinates) is eventually closed. Long-range CJT order is stable only in a relatively small region of the phase diagram (\emph{cf}.\ the manganites\cite{Lynn_2007}), corresponding to low vacancy concentrations and high fractions of JT-active ions. An important consequence of the crystal-field term is that it breaks the symmetry of the ground-state strain with respect to JT composition observed in vacancy-free systems, leading to greater residual strain in JT-rich compositions once vacancies are introduced.

\begin{figure}
\begin{center} 
\includegraphics{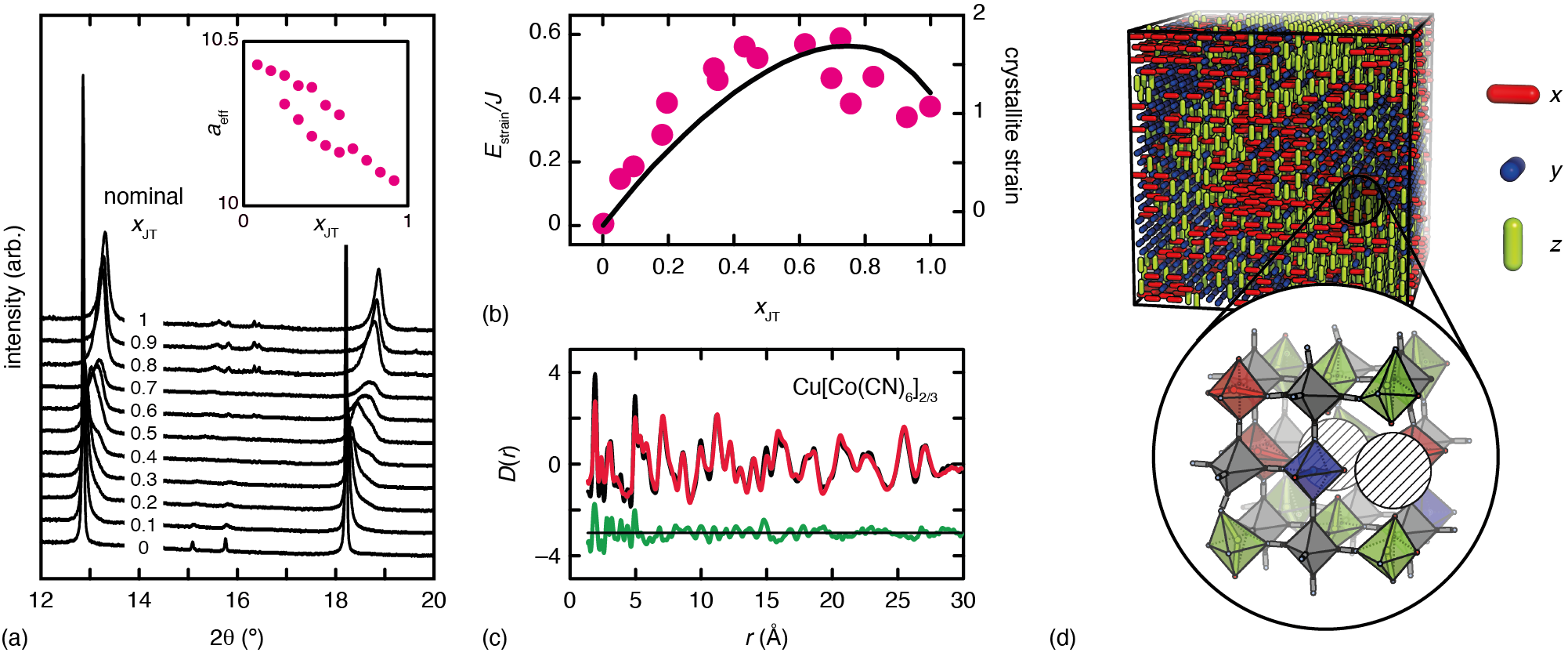}
\end{center}
\caption{\label{fig5}\footnotesize{\bf Structural behaviour of Mn$_{\mathbf 1\boldsymbol{-x}}$Cu$_{\boldsymbol x}$[Co(CN)$_{\mathbf 6}$] PBAs.}  (a) Representative regions of the X-ray diffraction patterns ($\lambda = 0.82507$\,\AA) measured as a function of M-site JT activity. The inset shows the variation in cubic lattice parameter as a function of $x_{\rm JT}$ determined using Pawley refinements. (b) The variation in isotropic crystal strain determined in these refinements (coloured data points) showed an asymmetric dependence on JT fraction that was consistent with expectation from our MC model (black line). (c) The X-ray PDF (black line) measured for a sample of Cu[Co(CN)$_6$]$_{2/3}$ could be well accounted for by an interaction-space refinement\cite{Harbourne_2024} based on the JT orientations determined by MC simulations (red line). The difference (data $-$ fit) is shown in green, shifted vertically by 2.5 units. (d) Representation of the JT axis orientations in one such MC model, with the different directions of axial elongation shown using rods of different colours. The complex texture that emerges reflects the tension between coalignment and vacancy-driven reorientations governing JT orientations. The inset shows a representative region drawn from a much larger interaction-space model used in the PDF fit, with Cu$^{2+}$ coordination polyhedra coloured according to the JT axis orientation and [Co(CN)$_6$]$^{3-}$ vacancies shown as hashed circles. Note the way in which vacancies affect JT orientations to give a local structure different to that found in CJT-ordered states [Fig.~\ref{fig1}(b)]. Standard uncertainties in the data shown in panels (a) and (b) are smaller than the symbols.}
\end{figure}

To test these general predictions, we studied the structural behaviour of a third and final PBA family of composition Mn$_{1-x}$Cu$_x$[Co(CN)$_6$]$_{2/3}$. These materials share a common hexacyanometallate vacancy fraction of $\frac{1}{3}$ but vary systematically in the degree of M-site JT activity. Our model anticipates that, in contrast to the miscibility gap of Mn$_{1-x}$Cu$_x$[Pt(CN)$_6$], this new family should exist as a solid solution. Our experimental X-ray diffraction measurements are summarised in Fig.~\ref{fig5}(a); we do indeed observe a series of solid-solutions with cubic $Fm\bar3m$ symmetry throughout this partially-vacant family albeit with some unusual peakshapes at intermediate compositions. These peakshapes were characteristic of a mixture of crystallites with slightly different compositions; suitable fits to these data could be obtained using a two-phase model, with the compositional difference between these two phases now relatively small. Note that, however these peaks are modelled, there are no longer any `forbidden' compositions in the series. Our Rietveld refinements allowed us once again to estimate the compositional dependence of isotropic strain, and we found excellent qualitative agreement between experiment and the residual (ground-state) strain anticipated from MC simulations---in particular we recovered the compositional asymmetry highlighted above [Fig.~\ref{fig5}(b)]. So the structural behaviour of this family is well captured by our simple strain/crystal-field model: both experiment and model demonstrate that vacancy incorporation acts to suppress CJT order and stabilise phases with stronger JT mixing.

Access to coarse-grained representations of JT orientations in JT-dilute and/or partially-vacant PBAs then allowed us for the first time to construct atomistic models of the local structure in these systems. We illustrate this point with the single representative example of Cu[Co(CN)$_6$]$_{2/3}$, chosen because it has at once both high JT activity and high vacancy content. To develop an atomistic model of the local structure in this system, we used the JT arrangements from the relevant MC simulation (\emph{i.e.}\ as carried out above) to decorate a suitable lattice-dynamical model as described in Ref.~\citenum{Harbourne_2024}. The model included coordinated water molecules for Cu$^{2+}$ ions adjacent to vacancies, but did not include zeolitic water. The model parameters (equilibrium bond lengths, effective force constants) were systematically varied in order to best fit the experimental X-ray pair distribution function (PDF), shown in Fig.~\ref{fig5}(c). This is a parameter-efficient approach for fitting the PDF that nonetheless gives a much better fit than is possible using conventional small-box Rietveld methods (here, $R =$ 30.6\% \emph{vs} 53.3\%, see SI).\cite{Harbourne_2024} A representative fragment of the local structure is illustrated in Fig.~\ref{fig5}(d), which captures how the presence of vacancies drives reorientation of the JT elongation axes to give a local structure distinct to that found in the fully-ordered state [\emph{cf}.\ Fig.~\ref{fig1}(a)]. This picture extends the phenomenology of orbital disordered states---which are notoriously difficult to study experimentally\cite{Thygesen_2017}---beyond that of dense phases such as perovskites\cite{Bozin_2007,Batnaran_2025} and transition-metal oxides.\cite{Browne_2017,NagleCocco_2024}

\section*{Concluding Remarks}

We have shown that cooperative JT order in PBAs is governed by the competition between strain and crystal-field considerations. The tension between these two effects leads to a distinction between long- and short-range CJT order in PBAs that is reminiscent of the complexity of orbital order in the manganite perovskites. In La$_{1-x}$Ca$_x$MnO$_3$, for example, the C-type Mn$^{3+}$ orbital order of La-rich compositions quickly melts on Ca doping to give a complex polaron liquid/glass phase from which the stripe charge/orbital order of La$_{0.5}$Ca$_{0.5}$MnO$_3$ eventually emerges.\cite{Lynn_2007} Of course, charge ordering is less relevant to the PBAs we have studied here, but the interplay between JT dilution and orbital orientations will nonetheless affect other electronic properties---\emph{e.g.}\ magnetism\cite{Tokoro_2004} or photoinduced charge-transfer processes\cite{Azzolina_2021}---in a conceptually similar way. The key microscopic distinction is that, in PBAs, the anisotropy competing with strain originates primarily from local crystal-field effects. While exchange mechanisms (\emph{e.g.}\ Kugel--Khomskii) may still be present on a much smaller energy scale, we have not needed to invoke these in order to explain our experimental measurements. An implication of this role of crystal field is that the use of different capping ligands in partially-vacant PBAs may provide a mechanism for (de)stabilising CJT order, \emph{e.g.}\ to avoid phase transitions during electrochemical cycling. Mechanically, JT disorder appears to strengthen PBAs relative to JT-inactive analogues,\cite{Bostrom_2024} which may be a consequence of nontrivial JT textures frustrating soft mode instabilities.\cite{Meekel_2024} We note for completeness that our analysis has not considered dynamical Jahn--Teller effects, which may be relevant irrespective of the presence or absence of CJT order.

The apparent insensitivity of our results to the presence and nature of vacancy correlations is perhaps surprising, but ultimately helpful. It is not yet clear, for example, whether the vacancy correlations known to occur in PBA single crystals\cite{Simonov_2020} are actually relevant to polycrystalline samples---such as those studied here or as found in functional PBA-based devices. The advent of electron-diffraction techniques for characterising correlated disorder in polycrystalline samples\cite{Schmidt_2023,Poppe_2024} may offer further experimental sensitivity to the nature of local CJT order in PBAs, and in this spirit we have included in our SI some calculations of diffuse scattering patterns based on the model developed in our study. Future studies may also consider the potential roles of A-site cations, which we have shown elsewhere to influence alternative distortion mechanisms.\cite{Cattermull_2023,Duyker_2016}

The interplay of strain and crystal-field effects that governs CJT order in PBAs is likely to be relevant also to similar phenomenology in other hybrid framework materials. In the formate perovskites, for example, CJT order can couple with cooperative tilting to break inversion symmetry,\cite{Stroppa_2011,Bostrom_2018} and so manipulation of local CJT order may allow the design of relaxor-like phases containing polar nanoregions.\cite{Evans_2016,Donlan_2017} Likewise in thiocyanates, where CJT order is crucial in stabilising low-dimensional quantum magnetism.\cite{Cliffe_2018} What differs amongst these various systems is the geometry of the lattice on which JT-active ions sit and the rules governing vacancy incorporation.\cite{Bostrom_2019b} Hence the calculation (and further validation) of ground-state phase maps for these various geometries of the kind shown in Fig.~\ref{fig4}  is an obvious priority in generalising the results of this study beyond PBAs. What is already clear is that molecular frameworks can support nontrivial CJT textures that (at least in the case of PBAs) can now be understood---and ultimately controlled---in terms of a small number of simple microscopic interactions.

\section*{Data Availability}
The datasets generated during and/or analysed during the current study are available from the corresponding author on reasonable request. Source Data are provided with this paper. 

\section*{Code Availability}
The Monte Carlo code used in this study will be made freely available in a public repository upon acceptance.

\begin{addendum}
 \item The authors gratefully acknowledge funding through the E.R.C. (Advanced Grant 788144 to A.L.G.), the U.K.R.I. (Frontier Research Grant EP/Z534031/1 to A.L.G.), the S.T.F.C. (studentship to E.A.H.), the Independent Research Fund Denmark (D.F.F.) (International Postdoctoral Grant 1025-00016B to N.R.), and the Royal Society through the Faraday Discovery Fellowships Fund, provided by DSIT. We also gratefully acknowledge the provision of beamtime at the Diamond Light Source: on beamlines I15-1 (experiment CY26330) and I11 (experiments II16997, EE13284, and CY25166-9). A.L.G. thanks Arkadiy Simonov for useful discussions, and Chloe Coates, Emily Reynolds, Arianna Minelli, Simon Cassidy and the beamline staff for assistance with the collection of diffraction data. 
 
\end{addendum}

\section*{References}

\bibliography{ncomms_2026_jtpbas}

\end{document}